\documentclass[reprint,amsmath,amssymb,aps,pra,superscriptaddress,showkeys,floatfix]{revtex4-1}

\usepackage{color}
\usepackage{graphicx}
\usepackage{dcolumn}
\usepackage{bm}
\usepackage[normalem]{ulem} 
\usepackage{float}

\begin{document}

\title{Spatially shaping waves to penetrate deep inside a forbidden gap}

\author{Ravitej Uppu}
\altaffiliation{Present address: Center for Hybrid Quantum Networks (Hy-Q), Niels Bohr Institute, University of Copenhagen, Blegdamsvej 17, 2100-DK Copenhagen, Denmark}
\author{Manashee Adhikary}
\author{Cornelis A. M. Harteveld}
 \author{Willem L. Vos}
\email{w.l.vos@utwente.nl}
\affiliation{Complex Photonic Systems (COPS), MESA+ Institute for Nanotechnology, University of Twente, P.O. Box 217, 7500 AE Enschede, The Netherlands}

\date{21 July 2020} 


\begin{abstract}
It is well known that waves incident upon a crystal are transported only over a limited distance - the Bragg length - before being reflected by Bragg interference. 
Here, we demonstrate how to send waves much deeper into crystals, by studying light in exemplary  two-dimensional silicon photonic crystals. 
By spatially shaping the optical wavefronts, we observe that the intensity of laterally scattered light, that probes the internal energy density, is enhanced at a tunable distance away from the front surface. 
The intensity is up to $100 \times$ enhanced compared to random wavefronts, and extends as far as $8 \times$ the Bragg length. 
Our novel steering of waves inside a forbidden gap exploits the transport channels induced by unavoidable deviations from perfect periodicity, here unavoidable fabrication deviations. 
\end{abstract}


\maketitle

Completely controlling wave transport is a key challenge that is essential for a large variety of applications. 
For instance, classical transport of acoustic waves has enabled sensing, ultrasound imaging and navigation~\cite{fink2000rpp, cummer2016nrm}.
In the quantum regime, control over electron and spin transport has led to major advances in the operation of nanoelectronic devices~\cite{neusser2009advmat,wagner2016natnano,Klyukin2018PRL}. 
In photonics, control over light transport has been exploited both in the classical and quantum domain, which has led to rapid advances in science and technology such as solar cells, quantum light sources, optical memories, and micro to nanoscale storage cavities~\cite{aspelmeyer2014rmp,Li2018opex,tandaechanurat2011natphot,obrien2009natphoton,kuramochi2014natphot,koenderink2015science}. 

\begin{figure}[htpb]
\centering
\includegraphics[width=\columnwidth]{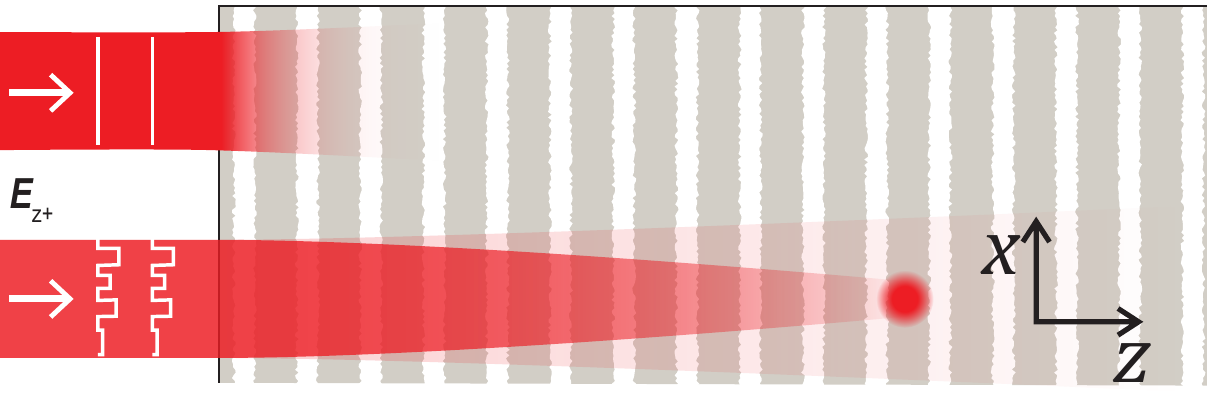}
\caption{
Schematic of wave transport in a two-dimensional photonic crystal consisting of pores in the x-direction that reveal unavoidable disorder such as roughness. 
}
\label{fig:fig1}
\end{figure}

An important tool in wave control is to exploit gap formation as a result of periodic and aperiodic symmetry.
The long range periodic order leads to a band structure in the dispersion relations of the waves. 
Forbidden frequency ranges, stop gaps, emerge in the band structure as a result of interference between the incident waves and Bragg diffracted waves~\cite{ashcroft1978book, joannopoulos2011book}. 
Bragg interference causes incident waves with a frequency in the stop gap to be exponentially attenuated, with a characteristic length scale called the Bragg length $L_B$, as is shown in Fig.~\ref{fig:fig1}.
The existence of gaps has led to exciting applications such as control of spontaneous emission, efficient light harvesting devices, and biosensing \cite{upping2012solar,demirci2017biosensor}. 

Real crystals feature unavoidable disorder resulting in broken symmetry due to, \textit{e.g.}, thermal motion and phonons in atomic crystals at finite temperature, quantum motion down to zero temperature, or unavoidable structural disorder in assembled photonic or phononic structures~\cite{koenderink2003prl}. 
The disorder gives rise to new channels for wave transport due to the multiple scattering, which are typically uncontrolled and thus detrimental for the applications mentioned above. 
In a specific realization of a crystal the microscopic configuration of disorder is fixed~\cite{grishina2019acsnano} and hence the disorder-induced channels do not in themselves offer new control. 
In disordered media without gaps, however, it is known that spatially shaping the phases of incident waves serves to set interferences between channels that represent new control: wave front shaping~\cite{mosk2012natphot,vellekoop2015opex,rotter2017rmp}. 

Therefore, in this paper we set out to demonstrate the tunable control of wave transport in real crystals. 
By spatially shaping the incident wavefronts as illustrated in Fig.~\ref{fig:fig1}, we steer the waves with frequencies within a gap to any desired location. 
The waves reach a remarkable depth of no less than $8 \times L_B$ with more than $30\times$ intensity enhancement. 
Remarkably, the shaped intensity at $5 \times L_B$ in the crystal is even $10 \times$ higher than the incident intensity $I_0$ in \textit{absence of wave shaping}, instead of being Bragg attenuated as is usual in periodic media. 
In our approach we employ wavefront shaping of light in photonic crystals that can be readily extended to electrons and acoustic waves, where analogous techniques have been demonstrated~\cite{harris2015structured,xie2014wavefront}. 

\begin{figure}[htpb]
\centering
\includegraphics[width=\columnwidth]{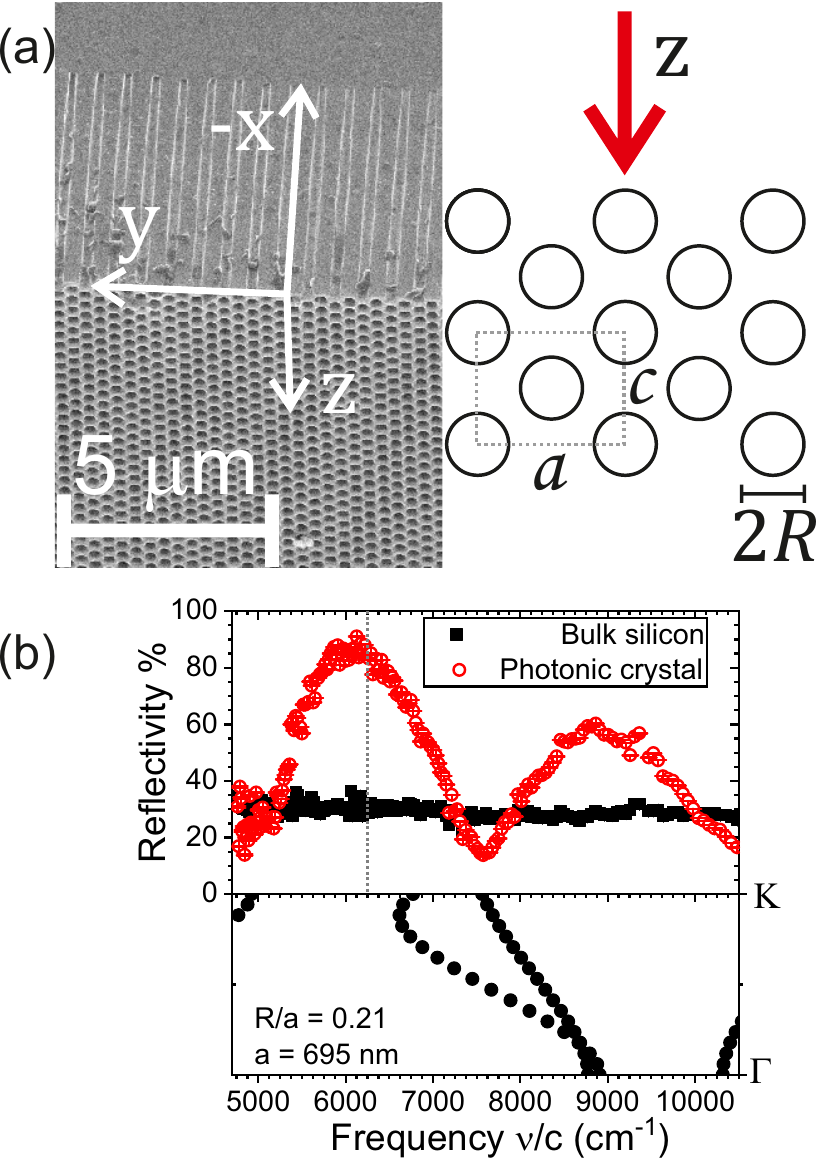}
\caption{(a) Left: scanning electron microscope image of the perspective view of a two-dimensional (2D) photonic crystal. 
The lower half shows the 2D array of pores in the $yz$-plane, and the upper half shows the $xy$ cleavage plane with an array of pores that extend up to $6~\mu$m into the silicon (see $5~\mu$m scale bar). 
Right: top view of the $yz$-plane showing the centered rectangular crystal with lattice parameters $a = 695$ nm and $c = a/\sqrt{2}$. 
The pores have a diameter $2R = 290$ nm (or $R/a = 0.21$).
Light is incident in the $z$-direction, corresponding to the $\Gamma K$ high-symmetry direction.
(b) Reflectivity measured on the photonic crystal (open circles) for TE-polarized light, compared to a reference of a silicon wafer (filled squares).
The prominent photonic crystal reflectivity peaks match well with the stop gaps in the calculated band structures. 
The dashed line highlights the frequency near the gap center where many steering experiments were performed. 
}
\label{fig:SEM_scheme}
\end{figure}
As exemplary waves, we study the propagation of light in two-dimensional (2D) photonic crystals that consist of large periodic arrays of pores etched in a silicon wafer~\cite{woldering2008nt,huisman2012prl}, see Figure~\ref{fig:SEM_scheme}(a). 
The pores are made by CMOS-compatible methods, employing deep reactive ion etching though an etch mask. 
The lateral $yz$-extent of the 2D crystal is $10\times$10 mm$^2$ at the center of the wafer, much larger than the expected Bragg and scattering length scales. 
The fabricated pores are about $6~\mu$m deep, sufficient for the focus to easily fit within the crystal. 
The wafers are cleaved in the $xy$-plane to expose the 2D photonic crystal to the incident light along the $\Gamma K$ high-symmetry direction~\cite{huisman2012prl}.

We developed a versatile experimental setup to perform optical wavefront shaping on silicon nanostructures, which operates in the near-infrared spectral range ($\lambda > 1100$ nm) where silicon absorption is avoided. 
The setup consists of three main components: 1) a broadband tunable coherent source, 2) a broadband wavefront shaper, and 3) a twin-arm imaging of reflected and lateral ($yz$-plane) scattered signal from the photonic crystal.
The broadband tunable coherent source was realized by spectrally filtering the emission from a supercontinuum source using a monochromator. 
A long pass filter (cut-on wavelength: 850 nm) is used to reject the background from second-order diffraction of shorter wavelengths. 
The available optical frequencies range from $4700$ to $11000$ cm$^{-1}$, corresponding to wavelengths $900 < \lambda < 2120$ nm, with a bandwidth of $0.6\pm0.1$ nm and a tuning precision better than $0.2$ nm. 
The filtered emission is collimated and expanded to a beam diameter of $7.5$ mm.

\begin{figure}[htpb]
\centering
\includegraphics[width=0.95\columnwidth]{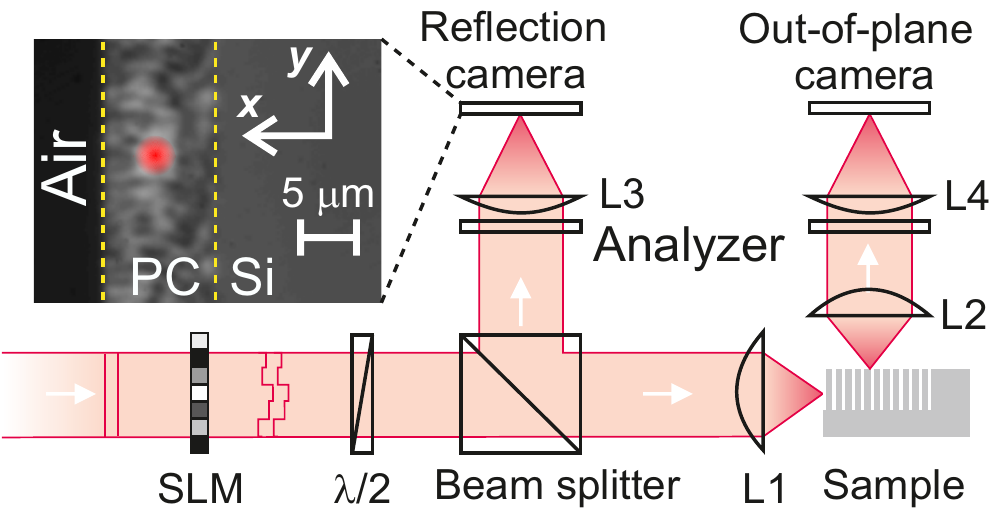}
\caption{
Linearly polarized light from a tunable source is directed off on a spatial light modulator (SLM) onto the sample through the microscope objective (M1).
The incident polarization on the sample is tuned using the half-wave plate ($\lambda$/2).
The reflected light from the sample is imaged onto a infrared camera using the same objective and a lens L1. 
A long working distance objective (M2) images the lateral scattered light onto a infrared camera using lens L2.
The lenses L1 and L2 have a focal length of +500 mm.
Broadband linear polarizers are used to analyze the reflected and the lateral scattered light.
The inset shows the image taken on the camera in the reflection arm.
The bright spot in the center (highlighted in red) is the focused coherent light with a constant phase of 0 rad displayed on the SLM.
}
\label{fig:setup}
\end{figure}

The wavefront of the collimated beam is then programmed employing a phase-only reflective spatial light modulator SLM (Meadowlark optics; AR coated: 850 -- 1650 nm), see Fig.~\ref{fig:setup}. 
The wavefront-programmed light is imaged to the back focal plane of the large numerical aperture ($NA=0.85$) infrared apochromatic objective lens $L1$.
The light reflected from the crystal is collected by the same objective.
The lateral scattered light from the $yz$-plane of the crystal is collected using a long working distance apochromatic objective $L2$ ($NA = 0.42$). 
Light collected by either $L1$ or $L2$ can be imaged to an InGaAs camera for aligning the incident beam. 
The inset in Fig.~\ref{fig:setup} shows the image captured in reflection, where the air, photonic crystal, and the unstructured silicon regions can be clearly separated. 
The bright spot in the center of the photonic crystal is the focused laser beam when a constant phase ($\phi = 0$) was programmed on the SLM. 

Reflectivity from the 2D crystal was measured following the procedures described in Ref.~\cite{adhikary2020opex}. TE-polarized reflectivity spectra shown in Fig. \ref{fig:SEM_scheme}(b) reveal two prominent peaks that are identified to correspond to two stops gaps in the calculated band structures.
We estimate the Bragg lengths for both gaps from the photonic strength $S$ using the relation $L_B = \lambda/(\pi S)$ \cite{vos2014book}, where the strength is defined to be $S \equiv \Delta\nu/\nu$ \cite{vos1996prb}. 
In the two stop bands at $\nu/c = 6250$ cm$^{-1}$ and $9000$ cm$^{-1}$, respectively, the photonic strength is $S = 0.25$ and $0.19$, respectively, corresponding to Bragg lengths $L_B = 2.0~\mu$m that are nearly the same for both stop gaps (see Fig.~\ref{fig:figa2} for details).

\begin{figure}[ht]
\centering
\includegraphics[width=\columnwidth]{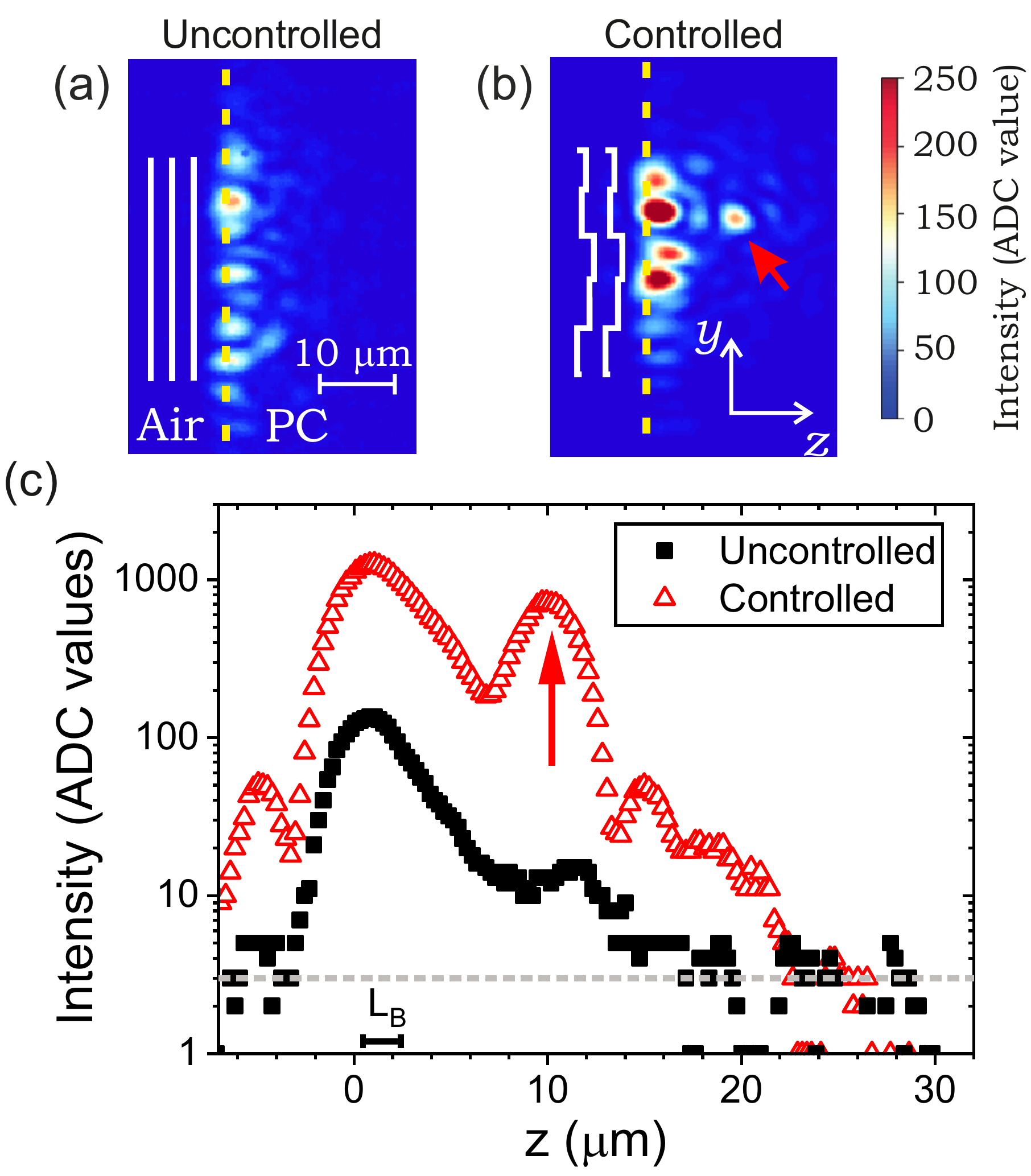}
\caption{
Camera images of the lateral scattered light ($\nu$/c = 6250 cm$^{-1}$) with (a) an uncontrolled and (b) the controlled wavefront. 
The red arrow points to the target location for steering the light, which exhibits as a bright spot. 
(c) The integrated intensity in a 5 pixel high ($yz$-plane) strip around the target location shows a clear increase in the intensity at the target depth of $10~\mu$m.
An overall enhancement in the intensity occurs due to the selective coupling of light into the deep-propagation modes of the crystal. The dashed grey line indicates the detector dark counts.
}
\label{fig:fig4}
\end{figure}

To demonstrate controlled transport of waves, we tune the frequency of the incident light to the center of the first stop gap at $\nu/c = 6250$ cm$^{-1}$ (i.e. $\lambda = 1600$ nm). 
The SLM was initialized with a phase pattern made up of 340 segments (each grouped from 32$\times$32 pixels) within the beam diameter, each assigned a random phase in the interval $[0, 2\pi)$.
The size of the segment was chosen to limit the illuminated area on the $xy$-plane of the sample to be confined within the photonic crystal.
The image of the lateral scattered light collected using $L2$ is shown in Fig. \ref{fig:fig4}(a), which shows a rapid decay of light into the crystal.
The input edge of the crystal is clearly discernible as the bright speckles (also marked by the dashed line).

We target a spot at a depth of $10~\mu$m - or $5 \cdot L_B$ - in the crystal and sequentially change the phase of each segment on the SLM to maximize the intensity at the chosen spot. 
Figure \ref{fig:fig4}(b) shows the image of the lateral scattered light at the end of the iteration over all the segments.
A distinct and bright focus is clearly observed at the targeted location, thereby demonstrating the first ever steering of waves inside a crystal, far beyond the Bragg length. 
Figure~\ref{fig:fig4}(c) shows the intensity around the targeted spot before and after steering the light, which shows $\approx100\times$ higher intensity with controlled transport.

In literature, it is common to characterize the quality of wavefront shaping by an enhancement $E_W$ that gauges the intensity increase at the target position~\cite{Vellekoop2008PRL, Popoff2014PRL}.  
The enhancement is commonly defined as $E_W \equiv I_{opt}/\langle I_r \rangle$, where $I_{opt}$ is the optimized intensity in the target spot and $\langle I_r \rangle$ is the intensity at the same spot that was ensemble-averaged over 100 random incident wavefront patterns. 
The intensity of the lateral scattered light is proportional to the energy density of the light that has scattered to a depth $z$ in the crystal. 
The enhancement at the target location inside the crystal is substantial, namely $E_W = 65 \pm 5$. 
Remarkably, the intensity at the target spot (depth $z = 5 \cdot L_B$) after steering is even $10\times$ greater than the intensity at the front surface ($z = 0$) \textit{before steering}, in contrast to traditional Bragg attenuation of waves inside any periodic medium. 
The enhancements observed here are a key highlight of the steering of the waves, which takes advantage of the multiple scattering of waves in the crystal due to the deviations from perfect periodicity~\cite{koenderink2005prb}. 

\begin{figure}[htpb]
\centering
\includegraphics[width=\columnwidth]{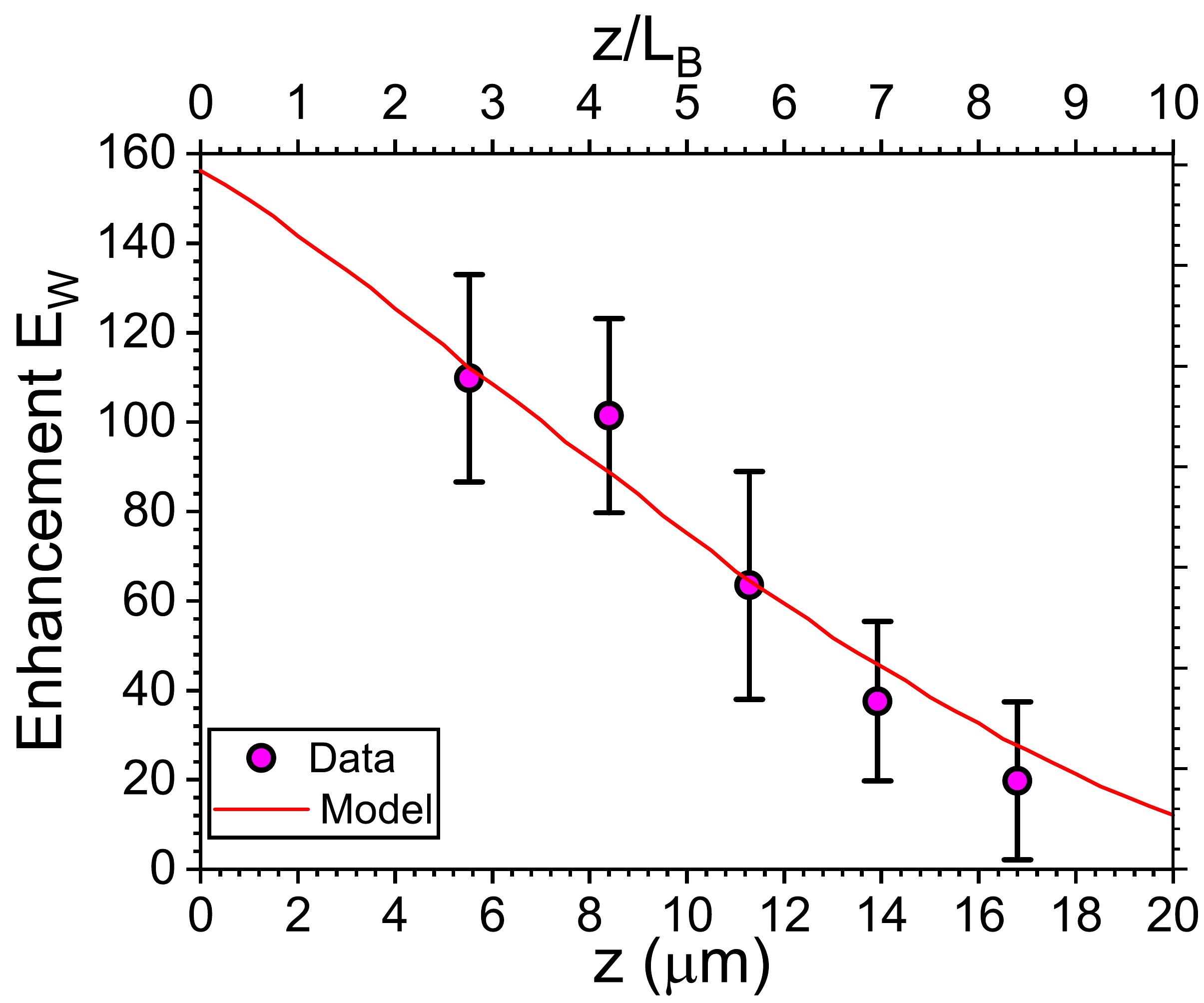}
\caption{
Measured wavefront shaping enhancement $E_W$ versus penetration depth $z$ into the crystal at a frequency at the center of the stop gap $\nu$/c = 6250 cm$^{-1}$. 
The calculated enhancement in a system with a synergistic interplay of the Bragg interference and random multiple scattering in a real photonic crystal agrees well with the measurements. 
[No. of modes (model) = 300; No. of modes (experiment) = 340]
}
\label{fig:fig5}
\end{figure}

Using wavefront shaping, we explore the maximum achievable depth inside the photonic crystal at which the intensity could be enhanced.
Figure~\ref{fig:fig5} shows the depth-dependent intensity enhancement $E_W$ at the center of the stop gap (at $\nu$/c = 6250 cm$^{-1}$).
The error bars represent standard deviations of $E_W$ measured at 8 different spatial locations along $y$ at the same depth $z$.
An enhancement $E_W$ in excess of 80 was achieved at depths up to $10~\mu$m, corresponding to $5 \cdot L_B$. 
At a depth of more than $8 \cdot L_B$, the measured enhancement is $E_W = 20$. 
We postulate that the decreasing enhancement $E_W$ with depth arises from the finite thickness of the photonic crystal that is limited by the pore depth. 
The finite thickness introduces surface losses that prevent the waves from reaching greater depths in the photonic crystal.

To interpret the intensity enhancement deep inside the crystal, we propose a new model that extends mesoscopic physics typical of random media~\cite{evers2008rmp, rotter2017rmp} to periodic crystals. 
The disorder-assisted multiple scattering of light inside the crystal is described using a transmission matrix $T$ over the $N$ transport channels.
The transmission matrix $T$ can be decomposed as $T = U D V^\dagger$, where $U$ and $V$ are unitary matrices of size $N\times N$ and $D$ is a diagonal matrix with values drawn from the DMPK distribution~\cite{dorokov1984ssc, mello1988annphys, evers2008rmp, akbulut2016pra}. 
The wavefront shaping enhancement $E_W$ is proportional to the largest eigenvalue of the matrix $T$, with the proportionality constant set by the wavefront shaping method~\cite{vellekoop2015opex}.
The presence of Bragg interference at frequencies within the stop gap and the surface losses result in the attenuation of the multiply scattered waves.
This attenuation results in a reduction of the number of transport channels with increasing depth inside the crystal.
Mathematically, the reduction in the number of transport channels is modeled as a truncation of the transmission matrix $T$, \textit{i.e.}, a reduced matrix size $M < N$ \cite{goetschy2013prl,hsu2015prl}.
The reduced number of channels $M$ with increasing depth in the crystal is modeled as $M(z) = N(1 - \alpha z)$, where $\alpha$ is an amplitude attenuation constant \cite{Pendry1990physica}.
The depth-dependent enhancement $E_W (z)$ is the maximum eigenvalue of the truncated transmission matrix. 
At a given $\alpha$ and $N$, we employ a numerical algorithm to generate an ensemble of random transmission matrices and compute the depth-dependent enhancement $E_W$.
The fit to the measured enhancement $E_W$ is shown in Fig.~\ref{fig:fig5} with $\alpha = (0.05 \pm 0.002)$ $\mu$m$^{-1}$ and $N = 300$, and is seen to agree very well with the data.
To put our model in perspective, the number of control parameters (segments) on the SLM is $N_{exp} = 340$, which agrees remarkably well with the number of channels in the model. 
The deviation could arise from the non-uniform intensity incident at each segment of the SLM due to the Gaussian profile of the incident beam.
From an independent measurement of the attenuation within the crystal, we extract the intensity extinction length $\ell_\textrm{ext} = 6.5 \pm 0.2$ $\mu$m (see Supplementary Material) arising from the multiple scattering of waves. 
The intensity attenuation length and the amplitude attenuation constant are related as $\ell_\textrm{ext} = 1/\sqrt{\alpha}$.
From the fit to $E_W$ in Fig.~\ref{fig:fig5}, we estimate $\ell_\textrm{ext} = 4.9 \pm 0.5$ $\mu$m, which is smaller than the measured value.
The underestimation of $\ell_\textrm{ext}$ is expected as the model effectively accounts for the attenuation due to the Bragg interference and multiple scattering of waves within the crystal, while the measured value is only due to the multiple scattering of waves.

In summary, we have demonstrated controllable enhancement of wave propagation at much greater depths than a Bragg length, even at frequencies within a band gap.
We take advantage of transport channels that are introduced in the crystal by unavoidable disorder, and address these by spatially shaping the wavefronts. 
The large depth to which waves are enhanced, even within a gap, broadens the range of applications feasible with photonic band gap crystals, both 2D and 3D. 
From the outset, photonic band gaps have been pursued for their radical control over spontaneous emission~\cite{lodahl2004nature, fujita2005science}, lasing~\cite{tandaechanurat2011natphot}, shielding of vacuum noise for qubits~\cite{Clerk2010RMP}, and for ultimate 3D waveguiding~\cite{Rinne2008NatPhot, Ishizaki2013NatPhot}. 
Based on our observations and modeling, light can be reconfigurably steered to resonant and functional features even inside a complete photonic band gap. 

Recently, combinations of gaps and tailored disorder are enjoying a fast-growing attention~\cite{Liew2011}, notably in phononic and photonic quasicrystals \cite{ledermann2006}, and hyperuniform \cite{man2013pnas,muller2017optica} and bio-mimetic structures~\cite{vignolini2018}. 
It is intriguing to speculate whether our crystals may reveal extremal transmission near a Dirac point and pseudo-diffuse behavior as predicted by Sepkhanov \textit{et al.} on 2D crystals ~\cite{sepkhanov2007pra}. 
Therefore our results open new avenues to increased wave control in many different classes of  metamaterials~\cite{yves2017natcomm,Ozawa2019rmp}. 

\appendix

\section{Spatial resolution of lateral scattered light} 
\begin{figure}
\centering
\includegraphics[width=\columnwidth]{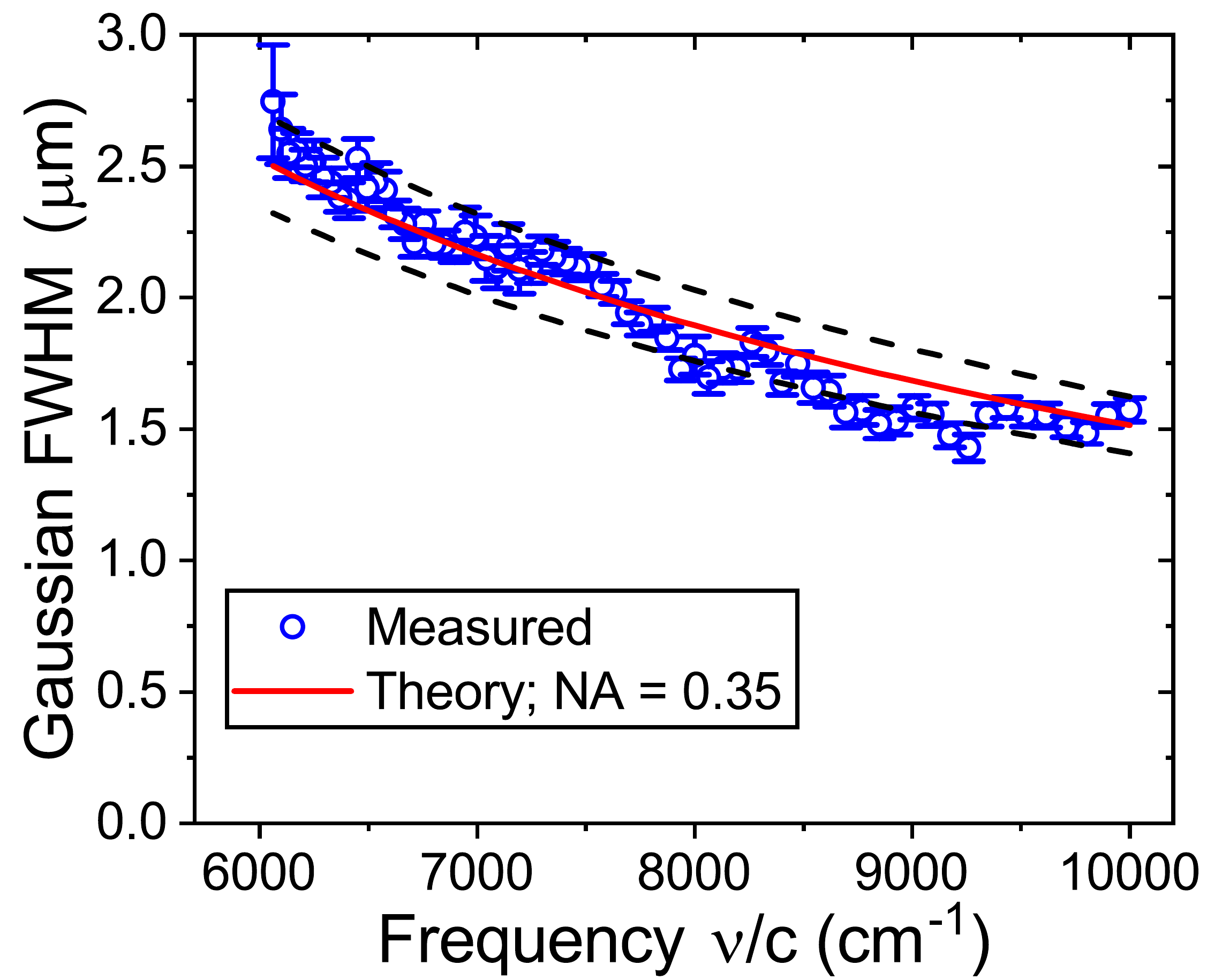}
\caption{The estimated Gaussian full width at half-maximum (FWHM) using the lateral-scattered images captured using the camera is shown versus input light frequency. 
We model the measurements to extract the collection aperture of the setup to be NA = 0.35.
The dashed curves are the 95\% confidence interval of the model. 
}
\label{fig:figa1}
\end{figure}

Structural disorder is known to result in a new length scale for wave transport, called extinction length $\ell_\textrm{ext}$ that statistically quantifies the strength of the disorder~\cite{akkermans2007book, wiersma2013natphot}.
The extinction length of light in the two-dimensional silicon photonic crystal was characterized by imaging the $yz-$plane of the sample on a InGaAs camera with an effective optical magnification of 125$\times$. 
A constant phase of 0 rad was displayed on the SLM to focus the light on the sample to a diffraction-limited spot.
The images of the lateral scattered light were captured at regular intervals (in $2$ nm wavelength steps) as the frequency of the incident light was varied from 6100 cm$^{-1}$ to 10000 cm$^{-1}$.
The intensity images were integrated along the height, corresponding to the $y-$axis. 
This depth-dependent intensity inside the crystal exhibits an exponential decay convolved with a Gaussian instrument response function, which determines the optical resolution.

Since the precise estimation of the resolution of the lateral scattering imaging setup is important in correctly estimating the extinction length, we first turn to this issue. 
The peak at the input edge of the photonic crystal was fit with a Gaussian to extract the resolution.
Figure~\ref{fig:figa1} shows the Gaussian full-width at half-maximum (FWHM) extracted from the fit.
We extract the numerical aperture for the collection arm to be NA = 0.35, which compares well with the nominal collection objective aperture NA = 0.42. 
It is reasonable that the effective aperture is slightly less than the nominal one, in view of some shadowing by the focusing objective at the input plane (the $xy-$plane). 

\begin{figure}
\centering
\includegraphics[width=\columnwidth]{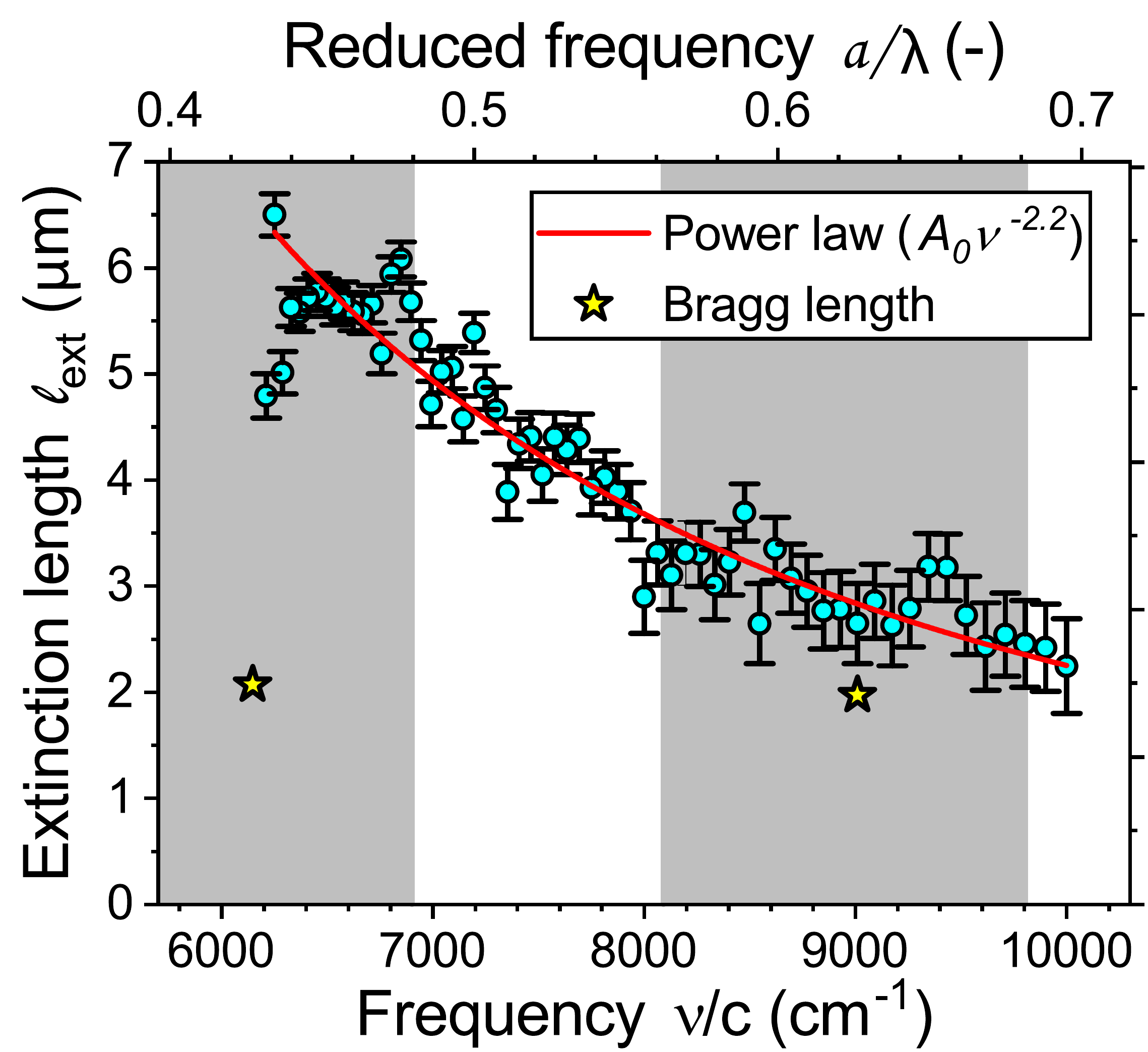}
\caption{The estimated extinction length $\ell$ of TE-polarized light propagating inside the 2D photonic crystal is plotted over a range of frequencies.
The frequency-dependence of $\ell$ closely follows earlier theoretical predictions of a power-law dependence with an exponent of -2.2.
The gray shaded regions correspond to the measured stop gaps.
The Bragg lengths $L_B$ as estimated from the photonic strength $S$ are marked as stars at the centers of the stop gaps.
}
\label{fig:figa2}
\end{figure}

\section{Extinction length of light in two-dimensional photonic crystals} 
Using the data from Fig.~\ref{fig:figa1}, we are now in a position to deconvolve the lateral scattered light with the corresponding Gaussian function. 
The deconvolved data were matched to a single exponential model to extract the extinction length $\ell_\textrm{ext}$.
Figure~\ref{fig:figa2} shows the extracted $\ell_\textrm{ext}$ (circles) as a function of frequency for TE polarized light incident on the photonic crystal. 
The error bars correspond to the 95\% confidence bound of the fitted extinction length. 
The extinction length decreases with increasing frequency, as expected, from about $\ell_\textrm{ext} = 6~\mu$m at the lowest frequencies in the first order stop band to about $\ell_\textrm{ext} = 2.5~\mu$m at the highest frequency  beyond the second stop band. 

To put these observations in perspective, we compare to theoretical work. 
Koenderink \textit{et al}. predicted a power-law dependence of the $\ell_\textrm{ext}$ on the frequency~\cite{koenderink2005prb}. 
For two-dimensional photonic crystals made of infinite long cylinders, the predicted dependence is $\ell_\textrm{ext} = A_0 \nu^{-2.2}$, where $A_0$ is a scaling parameter that depends on the degree and nature of the disorder.
We adjusted only the scaling parameter to the measured data in Fig.~\ref{fig:figa2} and observe that the power-law dependence on frequency agrees very well with the measurements. 
The observed deviation at low frequencies in the first stop gap is attributed to our choice of limiting the model to a single exponential to describe the intensity inside the crystal. 
Inside the stop gap, Bragg interference leads to additional extinction of light with the Bragg length $L_B$ as the characteristic length scale, depicted as star markers in Fig.~\ref{fig:figa2}. 
The two length scales would thus require a bi-exponential model to the intensity attenuation within the crystal, which is at this time difficult to significantly model, given the limited dynamic range and signal to noise ratio of the data. 
At frequencies in the range of the second stop gap, the Bragg length is close to the extinction length, and thus it is not sensible to try to fit the data with a bi-exponential model. 

\section{Wave front shaping of light in the photonic crystals} 
The error bars of the enhancements $E_W$ in Fig.~5 have nearly equal size, independent of the absolute magnitude of the enhancement. 
We surmise that the enhancement $E_W$ is Gaussian distributed, typical of independent observations, with a standard deviation given by the observed error bar. 
Conversely, the enhancement does not match with Poisson statistics, where the error bar would grow with magnitude.

\acknowledgments
We thank Diana Grishina, Ad Lagendijk, Willemijn Luiten, Femi Ojambati, and Allard Mosk (Utrecht) for helpful comments and experimental help. 
We acknowledge support from NWO-FOM-program``Stirring of light!'', STW-Perspectief program ``Free form scattering optics'', and MESA+ Institute section Applied Nanophotonics (ANP). 

%

\end{document}